Note: This article is a revised version of the report on Feb.7, which was originally written in Chinese. But it is more than an English translation of the prior report – we tried to revise it to make it more internationally relevant, also reflect the recent epidemic developments. The original report was attached intact.

# Visual Data Analysis and Simulation Prediction for COVID-19

Baoquan Chen[*], Mingyi Shi, Xingyu Ni, Liangwang Ruan, Hongda Jiang, Heyuan Yao, Mengdi Wang, Zhenghua Song, Qiang Zhou, Tong Ge

Center on Frontiers of Computing Studies, Peking University[#]
\* <baoquan@pku.edu.cn>   # <http://cfcs.pku.edu.cn/>

**Abstract**

The COVID-19 (formerly, 2019-nCoV) epidemic has become a global health emergency, as such, WHO declared PHEIC. China has taken the most hit since the outbreak of the virus, which could be dated as far back as late November by some experts. It was not until January 23rd that the Wuhan government finally recognized the severity of the epidemic and took a drastic measure to curtain the virus spread by closing down all transportation connecting the outside world. In this study, we seek to answer a few questions: How did the virus get spread from the epicenter Wuhan city to the rest of the country? To what extent did the measures, such as, city closure and community quarantine, help controlling the situation? More importantly, can we forecast any significant future development of the event had some of the conditions changed? By collecting and visualizing publicly available data, we first show patterns and characteristics of the epidemic development; we then employ a mathematical model of disease transmission dynamics to evaluate the effectiveness of some epidemic control measures, and more importantly, to offer a few tips on preventive measures.

## 1.Overview of the Epidemic Transmission

The heat map in Fig 1 visualizes the epidemic transmission in its early stage, based on daily diagnosed patient numbers across China. It's clear that the epidemic spread from Wuhan (Left: the hotspot in the middle) to its surrounding areas, and further to the rest of the country, especially the large metropolitan areas, such as Beijing, Shanghai, and Guangzhou (Right: hotspots from the North to South along the coast), which became secondary epidemic centers to its vicinities. Such transmission pattern reflects the large population movement prior to the Chinese New Year (January 25), when people usually return home to celebrate a week-long Spring Festival. On January 23, the Wuhan city exercised an extreme whole-city quarantine measure by shutting down all transportations going in and out of the Wuhan city.

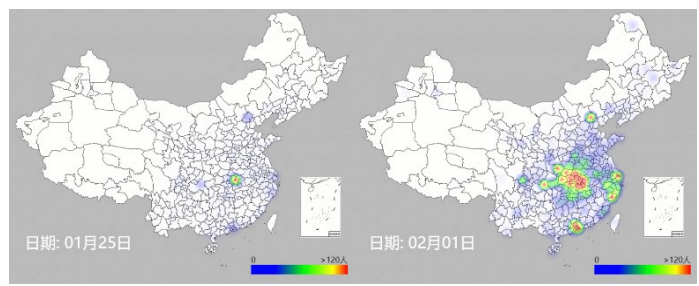

Figure 1. Heat map of epidemic spread in provinces and cities in China (L/R:Jan 25/Feb 1)

## 2. Visualization of Epidemic Spreading Pattern

We take the period January 31, approximately a week after the city closure, as the first stage of the epidemic development, after that, the second stage. The one-week duration is the averaged incubation time that it takes an infected person to show symptoms. At the first stage, the new Coronavirus infection cases diagnosed outside Wuhan are mainly imported from Wuhan through population movement. We collect data from several public sources [1][2] and visualize population volume on map.

From the visualization (Fig. 2), we can see strong correlation between the population of diagnosed infection and migration from Wuhan (Left-Middle) in the first stage, however, the epidemic development start to take different paces at different provinces. This reflects new cases of human-to-human transmission, with varying rates in different provinces. Next, we simulate epidemic dynamics to numerically analyze and predict the transmission procedures, and provide evaluation on different control measures.

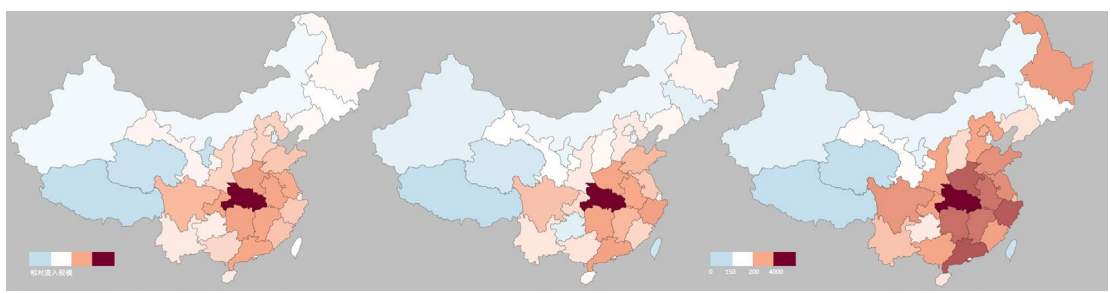

Figure 2. (Left) population flow from Wuhan to the rest of the country. (Middle) diagnosed infection cases in Jan. 31. (Right) diagnosed infection cases in Feb. 9.

# 3. Simulation of Epidemic Transmission Dynamics

On January 31, The Lancet published an article [3] on nowcasting and forecasting of COVID-19 disease (then called 2019-nCoV) spreading in China. In this work, the authors used the classic SEIR model, which divides the population into four categories of individuals: susceptible (S), exposed (E),infected (I), and removed (R). The model assumes that individuals transfer between categories with a certain probability. The regenerative number R0 is estimated to be 2.68. From this model, they estimated that the number of infected people in Wuhan had reached about 75,815 by January 25, and a daily infection number to be hundreds of thousands. They estimated several other populous cities (Beijing, Shanghai, Chongqing, Guangzhou) to have somewhat identical trend, except a couple of days shift on the peaking dates. The article basically dismissed the effect of the city-wide quarantine of Wuhan and other aggressive control measures around the country. The forecasting depicted a grave scenario both within China and globally, was helpful in alerting the public, but seemed to be quite off from the reality even at the time of its publishing.

One major problem of the above simulation is that it does not consider the aggressive control measures conducted around China, where (1) increasingly efficient testing procedures are put in place to timely diagnose infected individuals, who are then immediately quarantined, if not already; (2) individuals who had contact with the infected people are also quarantined; and (3) a further group of individuals who are suspected to be in the above two categories are also quarantined. The individuals in the latter two categories may be either released or moved to the first category after further observation and testing. In addition to these, there are more measures on reducing mobility within the city.

## 3.1 Our model

We aim to employ a new epidemic dynamics model to better consider the above measures in China. To this end, we employ the C-SEIR model [4], which has added two new quarantined groups in SEIR (Fig. 3.1): the quarantined suspected infection group (P), and the quarantined diagnosed infection group (Q). Individuals in groups P and Q do not have the ability to transmit the virus onward.

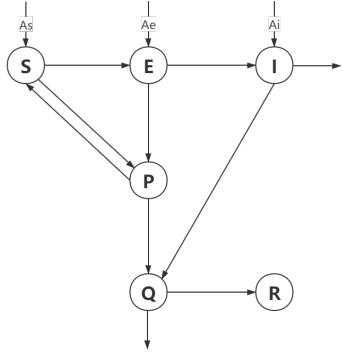

Figure 3.1. C-SEIR model

The virus infectious rate function β(t) is defined as the daily number of newly infected people ΔE divided by the number of untreated patients I. Considering that the new virus may also have a certain infectivity during the incubation period, it can be written as:

$$\beta(t) = \frac{\Delta E}{I + kE}$$

with k as 0.1, an empirical parameter meaning that the infectious capacity during the incubation period is 0.1 times that when symptoms are expressed. In order to determine the specific form of β(t), we first calculate the estimated value of β(t) using the daily number of confirmed patients and the estimated incubation period published by the Health and Medical Commission of China to infer the daily number of incubation periods and the number of infections, and then fit the data with certain chosen functions.

In order to compute the infectious rate β(t), it is necessary to calculate the number of exposed individuals (E) and the number of infected patients (I) for each day. Since only data of newly diagnosed patients are available, we need to estimate E and I based on it. Like in the article [4], we assume that the time between two generations of infection, and the time between infection and treatment of 2019-nCoV, are similar to SARS, which are 9 days and 3 days, respectively. That is to say, we can roughly assume that a patient exposed at day t will be infected at day t+6, and be treated at day t+9. Thus, it can be estimated that the total number of patients admitted during t ~ t + 9 is equal to the number of exposed individuals on day t, and the total number of patients admitted during t ~ t + 3 is equal to the number of infected patients on day t. From this, we can estimate E and I of each day, then β(t).

During the spread of most infectious diseases, the infectious rate β(t) decreases exponentially with time. We then use an exponential function to fit β(t) for each day. Taking Beijing as an example, the fitted result is shown in Figure 3.2. By applying similar calculation to the data from other provinces and cities, we have obtained largely similar infection rate curves. This means that the virus infection rates have dropped very quickly, a testimony to the effectiveness of the control measures.

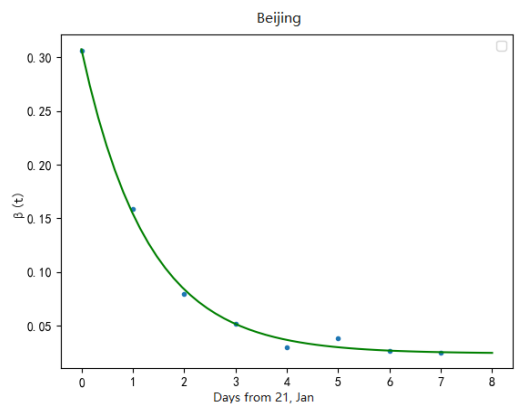

Figure 3.2. Fitted curve of infection rate change in Beijing

## 3.2 Simulation Prediction

We estimate model parameters and run the simulation, aiming to fit the output with the daily confirmed diagnose data reported from Hubei Province, Wuhan being its capital city. Through this, we obtain prediction on the future epidemic development, shown in Figure 3.3.

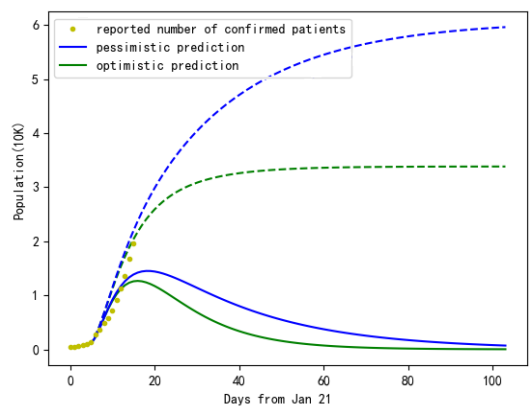

Figure 3.3. infection prediction (green and blue curves show results of two parameters)

Here, we generate two different prediction curves (blue / green), where the solid line is the number of daily confirmed patients and the dashed line the cumulative number of confirmed patients. In either case, the curve fits with the actual number of people diagnosed (beige point) in the early stage. It can be seen from the figure that slight deviation on this curve fitting in the early stage may lead to significant departure going forward. Nevertheless, the 'pessimistic' prediction (blue curve) gives a peak date around Feb 10, and the cumulatively confirmed patients number to be 50K on

March 1[st], which is quite close to the official numbers publicized. In stark contrast, the prediction of the Lancet article [2] is an order of magnitude that of the reality.

## 3.3 Evaluation of Control Measures

We like to conduct simulation by varying a few parameters to evaluate the effectiveness of different control measures.

First of all, proper quarantine is the most important, especially at the early stage of the epidemic development. It is believed that the outbreak of COVID-19 started as early as mid-November, but Wuhan city did not alert the public until Jan 20, and exercise effective quarantine measure until January 23[rd]. As can be seen from the simulation (Fig 3.4), shifting the quarantine date either earlier or later by 2 days could result in almost double amount of decrease or increase of the infected people cumulatively. It is important to note that too large a number of infected people may exhaust or even paralyze available medical resource. This unfortunately became the case for the Wuhan city, which was then rescued by pouring medical resources (medical personnel and supplies) from the rest of the country. The city also exercised an extremely aggressive whole-city quarantine on Jan 23, and a reduction of inter-city mobility to almost zero, with the aim of halting any disease transmission.

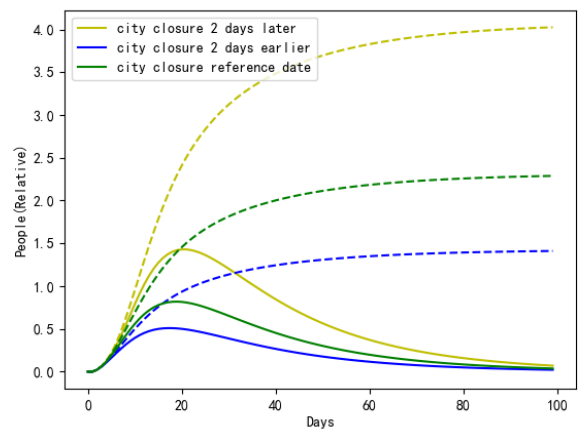

Figure 3.4.Simulation suggests that exercising quarantine at the early stage of the virus outbreak can lead to double decrease or increase of the total infections by shifting the quarantine date only 2 days earlier or later.

As of today (March 2, 2020),the epidemic in China is well under control. Outside Wuhan has seen zero to single digit daily confirmed infection for a week or so. This is achieved through the aforementioned aggressive quarantine measure across the nation, as described in the report by the WHO-China Joint Mission [6].

Next, we examine the impact of relaxing quarantine measures on epidemic development. Figure 3.5 shows predictions on different quarantine measures: if the quarantine is cancelled completely after reaching the daily peak, the rate of epidemic mitigation will be greatly reduced, resulting in a possible second peak of infection (blue), while a partially relaxed quarantine leads to only slight increase of infection (beige), comparing with the normal quarantine measure (green). This illustrates both the importance of quarantine to avoid a second strike, and the necessity of exercising a balanced measure to maintain life and work normalcy as much as possible.

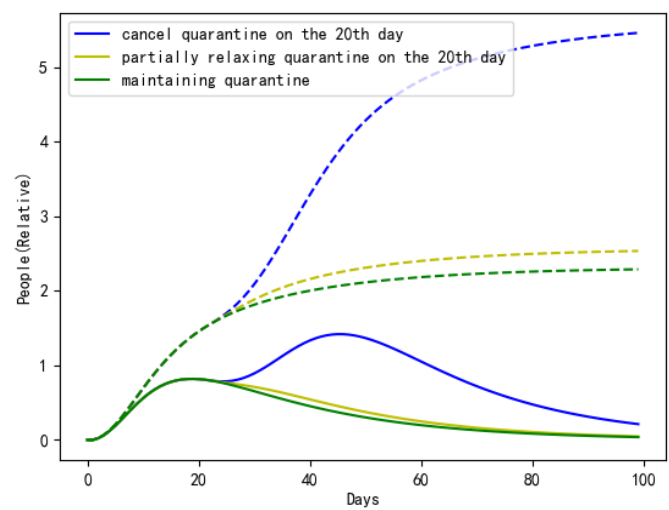

Figure 3.7. Prediction on different quarantine measures: cancelling quarantine completely (blue) results in a second peak, while a partially relaxed quarantine leads to only slight increase of the infected people (beige), comparing with the normal quarantine measures.

## 4. Concluding Remarks with Tips Offered

Our simulation based on the C-SEIR model has shown significant effectiveness of quarantine measures, which has been thoroughly exercised in China. While we applaud the success of these efforts, we must point out that this comes at a price much too high. In the hindsight, this could have been avoided by exercising a more proper (better balance between effectiveness and inconvenience) quarantine, but much earlier.

The COVID-19 virus seems to be spreading globally, every country has a different scenarios and government reactions. For individuals, there are a few points to consider:
(1) Be educated about the virus, e.g., its signature symptoms being fever and dry cough, unlike cold (running nose), and its prevention (not much different than preventing flu).

(2) Be vigilant and take extra precaution. Avoid big gatherings, stay low in social activities, and keep social distances in daily lives. Cancel or postpone non-essential travels.
(3) Quarantine if needed. Individuals with light symptoms, or having identified close contacts with those infected, should quarantine themselves in their own living places, or anywhere conditioned for the purpose. Individuals with strong symptoms should check in hospital.

An alert and corroborating public is probably the most important for containing the epidemic.

A recent article by the frontline health experts in China reported a death rate of 1.4% [7]. But outside Hubei province, the death rate seems to be much lower than this, down to 0.85%. The reported death rates outside China (e.g., U.S.) seem to be even lower. It is worth investigating whether the transmissibility of the virus would also reduce over time. Nevertheless, the COVID-19 virus is definitely more than a strong flu; therefore, it deserves extra precaution from individuals and well prepared medical resource from governments.

## Acknowledgment

We thank Professor Juan Zhang, author of the C-SEIR model [4], for sharing her insightful knowledge of the model since our prior report.

## References
[1] 2019-nCoV data repository: https://github.com/BlankerL/DXY-2019-nCoV-Data
[2] Baidu Map – migration: http://qianxi.baidu.com/
[3] Joseph Wu, Kathy Leung, Gabriel Leung. Nowcasting and forecasting the potential domestic and international spread of the 2019-nCoV outbreak originating in Wuhan, China: a modelling study. The Lancet, January 31, 2020.
[4] Zhang J , Lou J , Ma Z , et al. A compartmental model for the analysis of SARS transmission patterns and outbreak control measures in China. Applied Mathematics and Computation, 2005, 162(2):909-924.
[5] Jon Cohen, Scientists are racing to model the next moves of a coronavirus that's still hard to predict. Science, Feb. 7, 2020
[6] Report of the WHO-China Joint Mission on Coronavirus Disease 2019 (COVID-19) https://www.who.int/docs/default-source/coronaviruse/who-china-joint-mission-on-covid-19-final-report.pdf
[7] Guan et al,Clinical Characteristics of Coronavirus Disease 2019 in China. The new England Journal of Medicine, February 28, 2020. DOI: 10.1056/NEJMoa2002032; https://www.nejm.org/doi/full/10.1056/NEJMoa2002032

# 面向新冠疫情的数据可视化分析与模拟预测

陈宝权 北京大学前沿计算研究中心 baoquan@pku.edu.cn

模拟仿真：倪星宇、阮良旺、姚贺源、王梦迪; 数据可视化：史明镒、蒋鸿达、宋振华、周强、葛彤

## 目录

## 导言

  2019 年在武汉爆发的新型冠状病毒肺炎（国家卫健委简称 NCP）传播迅猛，已被世界卫生组织（WHO）定为"国际关注的突发公共卫生事件"。对疫情的控制，自 1 月 24 日武汉宣布封城之后，各个省市也陆续通过启动重大突发公共卫生事件一级响应来控制人口流动；同时，各省市医疗队伍驰援武汉，武汉的防控措施也急速加强；但全国疫情，特别是湖北省的状况依然让人揪心。公众非常关心疫情的发展趋势，期待"拐点"的出现；疫情防控部门希望不断总结经验教训，评估现有措施的有效性。该疫情的发展成为了涉及我国政治经济民生的一件大事。

  此次病毒的传播到底如何从武汉向外传播？不同省市疫情的发展呈现怎样的差别？封城、社区化隔离等一系列措施对减缓疾病传播起到了多大的作用？更为重要的是，拐点何时出现？我们的报告首先从已有数据的可视化来展示疫情传播特点，然后通过建立传染病动力学模型，评估疫情防控措施，提出建议并预警，同时预测疫情疾病走势，给疫情防控决策和大众行为提供参考。

## 1. 疫情传播可视化总览

  通过热度图的方式，我们使用国家及各省市地区卫健委公布的地级市每日确诊数据[1]，在图 1.1 中重现了 NCP 疫情的传播。容易发现，**疫情的传播主要**

以武汉为中心向周围扩散，通过人口流动将病情传播至中心城市，北京、上海、广州等地，成为二级传播中心。

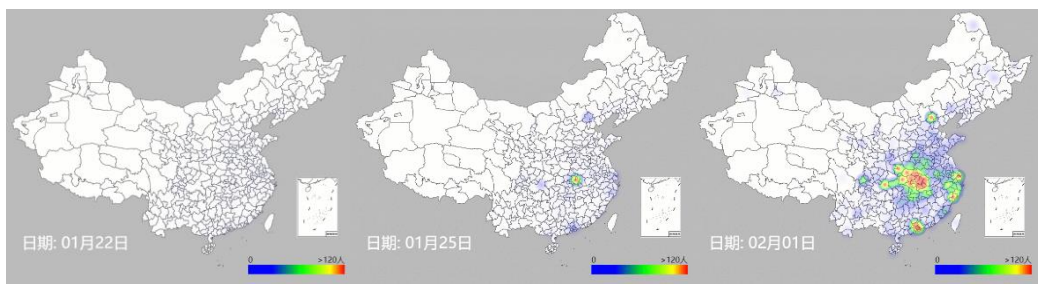

图 1.1. 各省市疫情传播热度图（1月21日至2月9日）

我们对全国、湖北及湖北以外的省市的新增感染人数可视化，容易发现，湖北以外各省，在 1 月 31 日前新增仍在不断增长，然后增速放缓，从 2 月 4 日开始有下降的趋势。而**湖北省的新增人数仍没有明显的下降趋势，加上检测的瓶颈，报告的数据可能和实际的情况相比存在较大的噪声，疫情防控形势依然严峻**。即使是湖北之外的其它省市，情况也各不相同，有些省市的情况亦不容小视，后面会展开分析。

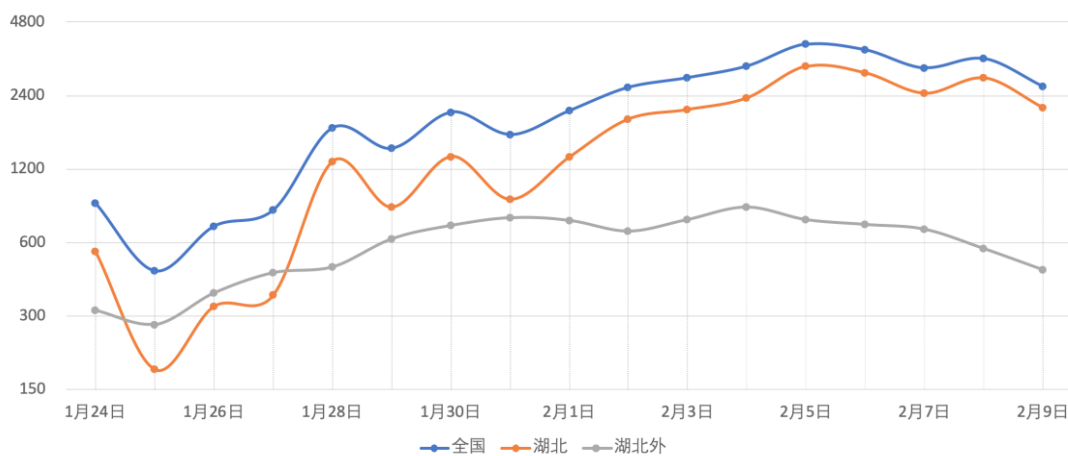

图 1.2. 湖北及湖北以外省市新增确诊人数的变化（1月24日至2月9日）

## 2．疫情传播特点分析

### 人口流动与疫情的不同阶段

人口流动是疫情发展第一阶段输入型传染的主要因素。为了具体描述其影响，我们使用百度迁移所提供的人口流动数据[2]，通过可视化春运期间从武汉流向全国各省市的人口规模(不包含港澳台数据)和全国感染病毒人数的分布，直观地观察两者间的联系。

疫情由湖北武汉华南海鲜市场开始传播，逐渐蔓延至全国。中国大陆各省份的颜色，反映了该省的确诊人数及来自武汉市的输入人流量。

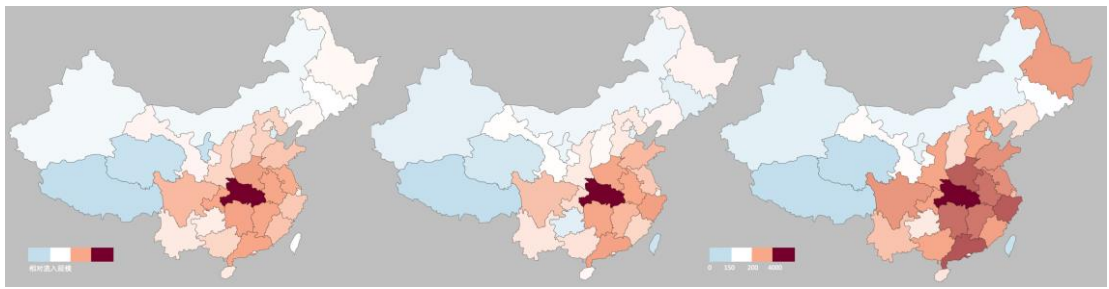

图 2.1. (左) 由武汉市流向各省市的输入人流量，(中) 1 月 31 日各省市确诊感染总人数，(右) 2 月 9 日各省市确诊感染总人数

通过对比图 2.1(左)和图 2.1(中)，我们可以看到，**在疫情初期，各省市感染总人数与春运期间由武汉市的输入人流量呈现强相关性**；需要指出的是，武汉 1 月 24 日封城，考虑平均潜伏期 7 天，1 月 31 日湖北外省市的确诊人群应该基本为输入型感染。但随着时间的推移，确诊人数分布图则发生了一定的变化（2.1(右)）。我们推断，武汉封城之后，二次传染所造成的病毒传播越来越占主导地位，和各省市的人口密度，以及管控措施等密切相关。

## 各省市传播差异

为了更具体分析各省市之间的疫情传播差异，首先，我们针对湖北以外的省市，以从武汉输入人流规模为基准，与当地截止到 2 月 9 日的确诊人数进行对比。见图 2.2：

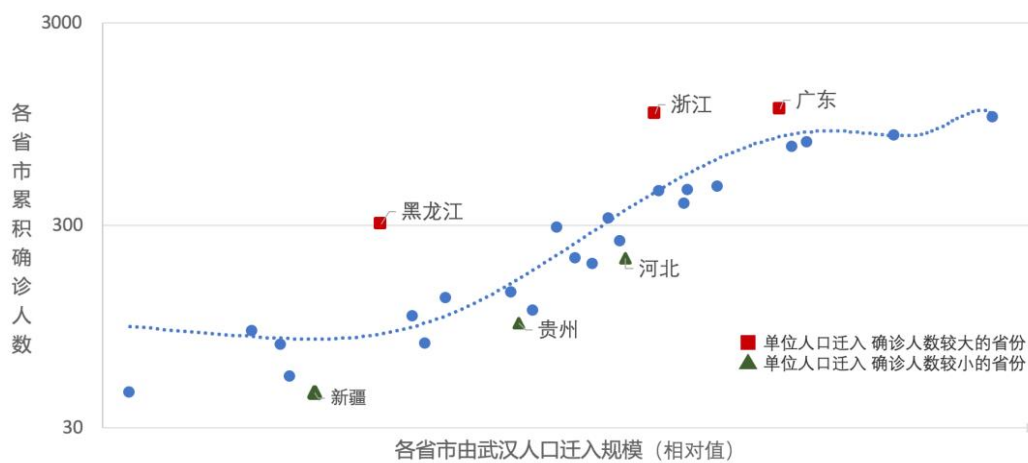

图 2.2. 湖北以外省市的武汉人口流入规模(相对值)与其确诊感染人数

从图 2.2 可以发现，**各省市的武汉输入人流量规模与其确诊人数之间存在正相关**，如图 2.1（中）一样，验证了人口流动是疫情初期传播的主要原因之一。然而，有些异常值出现，代表了疫情传播比较特殊的几个省份。

为了更好地观察这些差异，我们对确诊人数做数据归一化，将每个省市确诊总人数分别除以武汉输入人数规模和该省市总人口，得到两个曲线，见图 2.3：

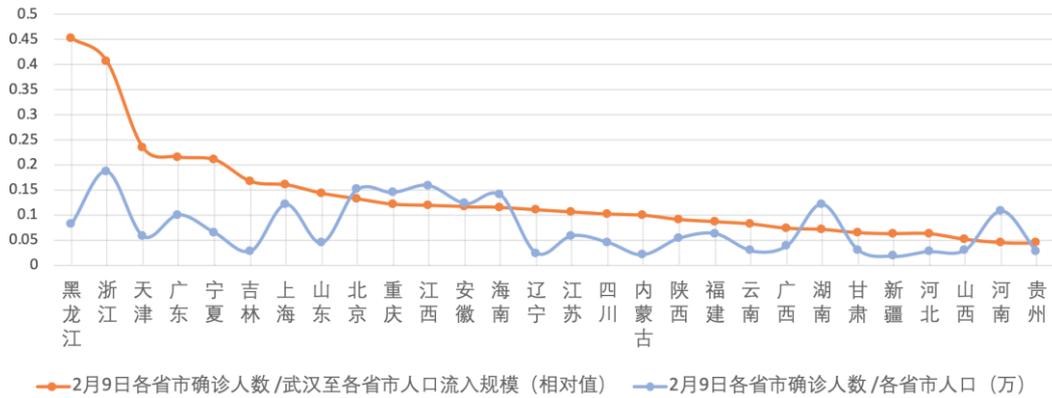

图 2.3. 截止 2 月 9 日，各省市确诊人数分别除以武汉输入人数规模以及该省市总人口数

对于单人感染率较高的省市（图 2.3 蓝色曲线），如浙江省、北京市、上海市、广东省，虽然确诊人数不是最多，但是发病率却相对较高，原因是这些省市都属于商务旅游集中、人流密集、流动性大的城市，因此造成传染性高于其他地区。有些省份公布了二次传播人数，黑龙江省的二次传染比例最高，有报道表示，至 2 月 6 日，黑龙江发生 48 起聚集性疫情传播，共导致发病 193 例。

**二次传播在疫情传播第二阶段中占主导地位**，对于目前聚集性传播的高发地区，采用更严格的隔离措施避免疫情的爆发型增长，是行之有效的方法。但后续疫情传播的走向具体如何，哪些因素更为关键，我们接下来采用传染病传播模型来做数字模拟和分析。

# 3. 疫情传播模拟

## 基础的 SEIR 模型

1 月 31 日，国际知名医学期刊《柳叶刀》发表了中国香港科学家的工作[3]。在该文中，作者采用了传染病动力学中经典的 SEIR 模型来进行模拟。该模型将人群分为易感人群（Susceptible）、已被感染但无症状处于潜伏期的人群（Exposed）、已表现出症状但未被隔离的患病人群（Infectious）、康复人群（Recovered）四类（模型把死亡人数也归到 R 中）。并假设他们之间按一定概率转移。其状态转移图如下：

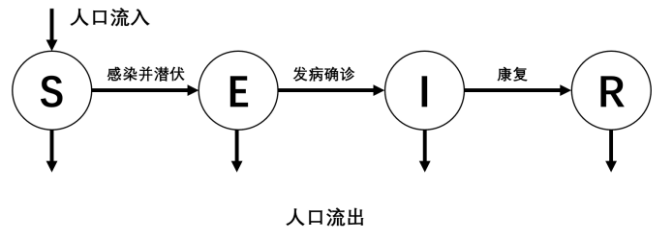

图 3.1. SEIR 传染病动力模型

该模型所涉及的参数主要为：可再生数 R0、平均潜伏期时间 DE 和平均收治时间 DI。其中，后两种参数均可直接从官方发布中获得，而 R0，即一名被感染者平均每天传染到的人数，其值较难估计。文章使用了 2019 年 12 月 31 日至 2020 年 1 月 28 日的感染人数数据，并根据境外（除香港）受感染人数及国际航班从武汉出境人数反推得 R0 为 2.68，采用如上模型推算出截止 1 月 25 日，武汉地区受感染人数约为 75815 人，预测疫情的拐点将在五月到来，并得出封城手段的采取对加快疫情缓解的作用不显著的判断。我们的观察是，该项工作的模型与参数选取存在不合理性，主要是：

1. 境外确诊数据样本量较小，且使用飞机这一交通工具的人群在总人口中并非均匀分布，据此假设泊松过程来估计 R0 偏差较大；
2. 考虑到政府防控措施的实施与升级，R0 的取值不应设为定值。尽管论文中假设戴口罩可以使 R0 减半并进行了一定的讨论，但这样的设置依然较为粗糙。
3. 封城作为非常严厉的防控手段执行得非常彻底，必须在参数设置中有效的反映。
4. 社区隔离措施作为后来使用的控疫手段也必须考虑。

最后一点（社区隔离）在现有的 SEIR 模型中无法模拟，为此，我们引入 C-SEIR 模型。

## C-SEIR 模型及模拟分析

C-SEIR 模型[4]相比于 SEIR 模型主要有以下两点改进：
1. 考虑政府的隔离措施，将人群进一步划分出隔离患者和未隔离患者，隔离患者不具备传播能力；
2. 考虑政府措施的加强和群众防护意识上升，病毒的基本可再生系数（R0）应该随时间变化而不是一个固定值，因此通过真实数据拟合出病毒的传染率曲线代替 R0。

针对第一点，C-SEIR 在 SEIR 的四类人群基础上增加了两类新的人群：被隔离疑似感染人群（P），已确诊并被隔离的患病人群（Q）。注意在 P 类中的人包括新冠状病毒的患者，也包括了症状相似但未感染新型肺炎的人群，可以假

设这一部分人群不具备向外传染病毒的能力，即病毒的传染能力只与 I 和 E 有关。同时，在考虑湖北省之外的省市时还需要考虑来自武汉的人群输入。

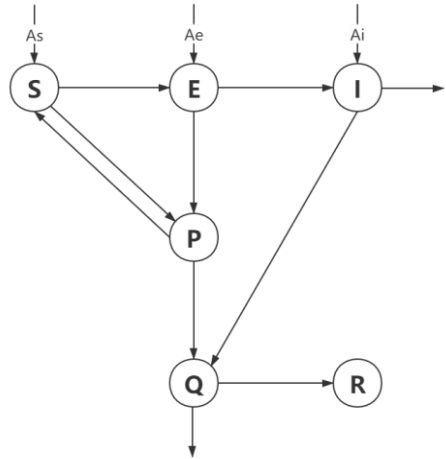

图 3.2. C-SEIR 传染病动力模型

对于第二点，首先定义病毒的传染率函数$\beta(t)$为每日新增的被传染人数 $\Delta E$ 除以未被收治的患病人数 I，考虑到新冠病毒在潜伏期可能也有一定的传染能力，可以写成公式为：

$$\beta(t) = \frac{\Delta E}{I + kE}$$

其中 k 取 0.1，表示潜伏期传染能力是表现症状时的 0.1 倍。为了确定$\beta(t)$的具体形式,我们首先使用卫健委公布的每日确诊人数以及估算的潜伏期时长反推每日的潜伏期人数和感染人数来计算$\beta(t)$的估计值，再选取函数对数据进行拟合。

为了估计$\beta(t)$，需要计算每天的感染人数 E 和发病人数 I。由于只能获得新增确诊人数的数据，所以需要以此为基础，对 E 和 I 进行估计。我们按照论文[4]中的方法，假定病毒的传代期和收治期和 SARS 相近，分别为 9 天和 3 天，也就是说，可以大致认为，第 t 天感染的人会在第 t+6 天发病，第 t+9 天被收治。由此即可估计：t~t+9 期间的总收治人数等于第 t 天的总感染人数，t~t+3 期间的总收治人数等于第 t 天的总发病人数，从而计算出$\beta(t)$公式中的各项，进而估计出$\beta(t)$。

在大部分传染病的传播过程中，传染率$\beta(t)$会随时间指数衰减。我们使用指数函数，拟合逐日估计的$\beta(t)$散点值。以北京市为例，拟合结果如下：

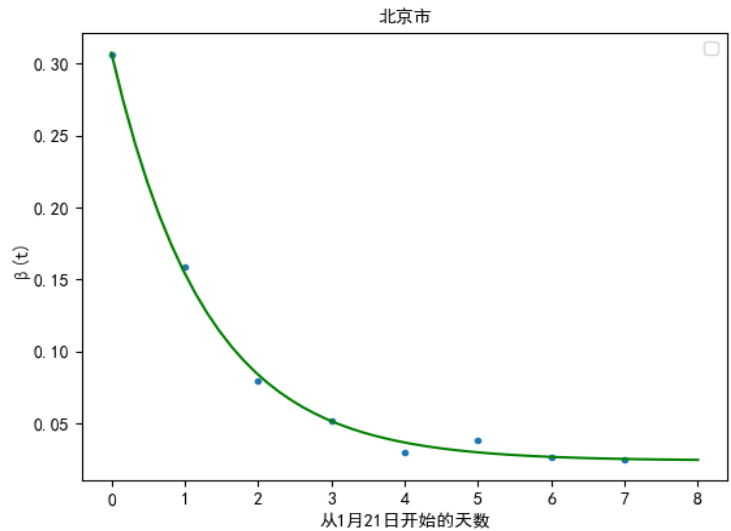

图 3.3. 北京市的感染率变化拟合曲线

可见，指数衰减的假设基本符合实际情况，能较好地描绘传染率的变化。我们将全国各地的曲线画在一张图上，到 2 月 7 日 24 时，累积感染人数超过 200 的各省份的拟合结果如图所示：

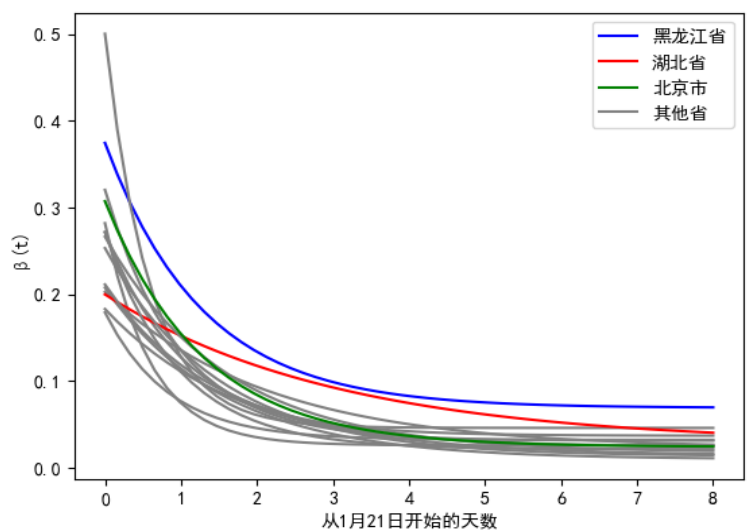

图 3.4. 疫情相对严重的省市的感染率变化曲线

从图中可以看出，**各省病毒传染率的下降均非常快，反映了控疫措施的成效**。其中，黑龙江与湖北两省的曲线离群，黑龙江的感染率收敛值偏高，而湖北的感染率下降速率偏慢。黑龙江的情况可参考前面的分析；湖北作为疫情源头，其感染率下降速度反映了该地区医疗资源的制约，控疫任务的艰巨性。

C-SEIR 模型中的其他参数依赖于病毒的特性，如潜伏期长度、病程、死亡率、治愈率，还依赖于政府措施的实施情况，如隔离人员的数量、发病到确诊所用的时间。我们使用与论文[4]类似的方法进行模型参数的确定，对湖北省确诊人数变化的数据进行拟合，并预测其未来的发展，如图 3.5 所示。

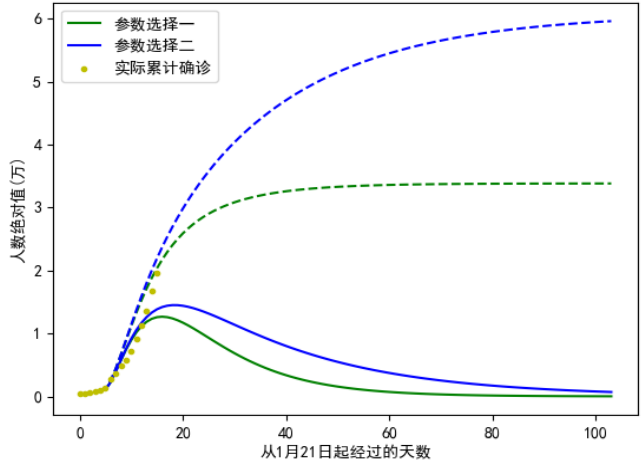

图 3.5. C-SEIR 模型预测曲线（基于湖北省确诊数据拟合）

　　这里，我们作出了两条不同的预测曲线（蓝色/绿色），其中实线为当天确诊人数，虚线为累计确诊人数。从图中可以看出，尽管两种参数选择在前期都与实际确诊人数（米黄色点）近似曲线吻合，并且拐点时间的预测非常接近，但是最终累计感染人数的预测相差非常大。这反映了一个事实：**在疫情初期尝试对疫情走向进行预测往往十分不准确，不能因为模型的预测而过分乐观或恐慌。**

　　考虑到各地疫情的新增确诊人数慢慢出现了拐点，我们也以北京市为例对新增确诊人数进行了拟合，如图 3.6 所示，采用了乐观（绿色）和保守（蓝线）两组参数来预测。首先新增确诊人数波动较大，因为考虑到上报的延迟可能导致新增确诊病例出现聚集，平均来看可以看到新增确诊病例确实有下降趋势。值得注意的是，新增病例出现拐点并不意味着疫情会马上消退，累计确诊依然会保持增长趋势，疫情的真正缓和由新增确诊的长尾来决定。**随着生产活动的逐渐恢复，广大民众应该依然保持防护意识，不能掉以轻心。**

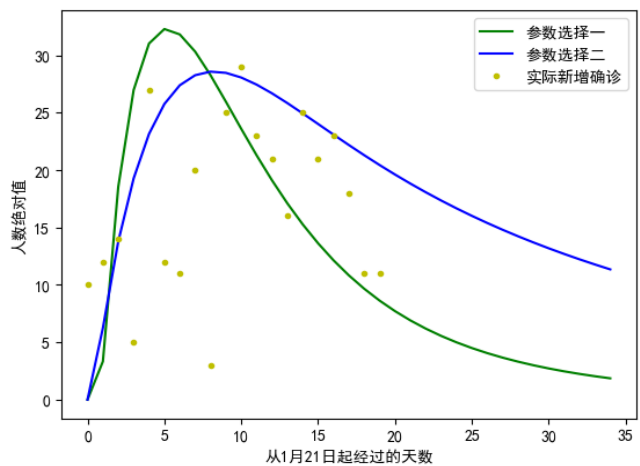

图 3.6. C-SEIR 模型预测曲线（据北京市新增确诊拟合）

　　同 $\beta$ 一样，模型中的其它参数也可能在不停地变化，并且极易受到突发事件的影响。鉴于以上因素，利用该模型预测疫情峰值的具体日期只有参考价值。

但是，通过半定量地分析具体参数，依然能够为今后政府的防控和个人行为提供参考。

首先我们考察隔离措施的持续对疫情变化的影响。图 3.7 对比了当确诊人数达到峰值后，是否立即取消对密切接触者的隔离对疫情的变化的不同效果。从图中可以看出，如果立即取消隔离，会大大降低疫情缓解的速度，甚至出现第二个峰值，因此，**保持高压防控不动摇，是接下来疫情防控的重中之重。**

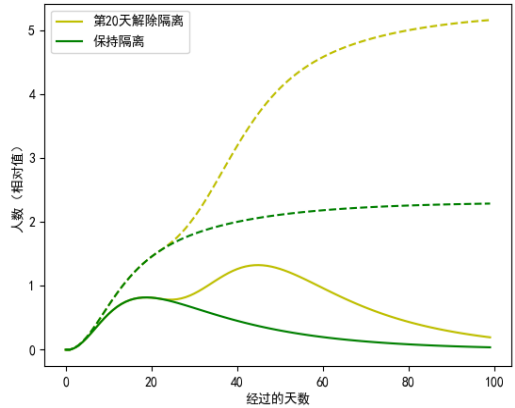

图 3.7. C-SEIR 模型预测对比：当疫情达到拐点后是否取消隔离

其次，每日新增被感染者中，已经被隔离者所占的比例也具有重要的意义。我们动态地调整该值，如图 3.8 所示。从该图中我们看出，尽管对疫情峰值时间的影响并不显著，但隔离比例降低会使得累积患病的人数成倍的增长。因此，**为了更快控制疫情，我们需要保持积极响应与配合各种有效的隔离措施。**

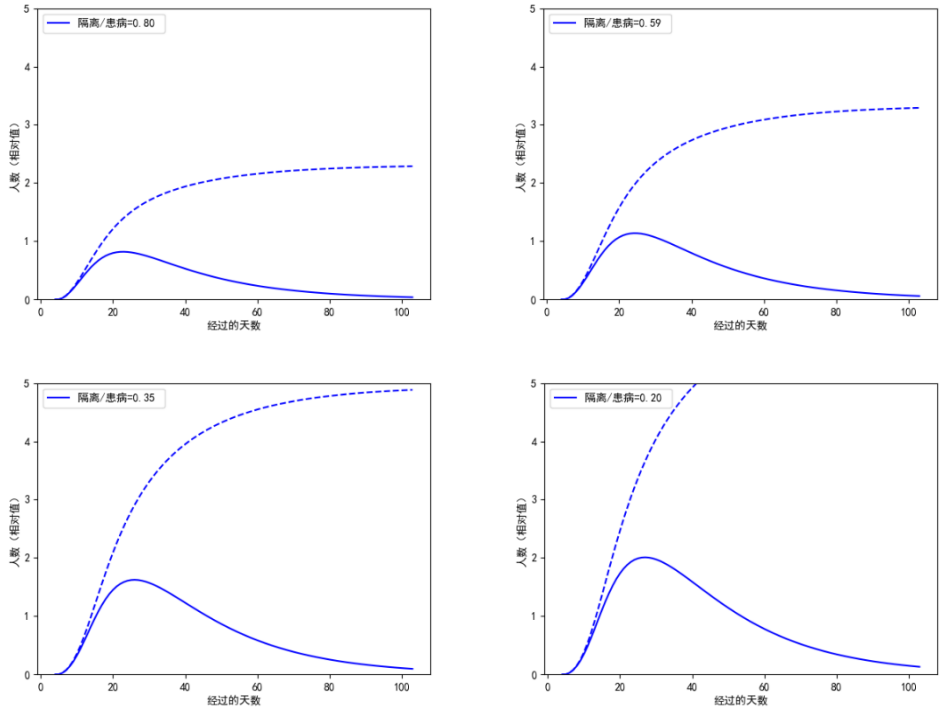

图 3.8. C-SEIR 模型预测对比：已隔离人员占患病者的不同比例

最后，我们试图复盘武汉封城这一控疫措施的有效性。我们利用该模型分析了武汉采取封城措施的时间点对疫情变化的影响，参见图 3.9。从中不难看出，在采取隔离等办法的情形下，**封城的提前或推迟不会对疫情拐点的到来时刻产生大的影响，但却会造成感染确诊人数的大幅度变化**。考虑到现实生活中有限的医疗资源，尽早地实施封城的措施是很有必要的。

今天的《Science》报道[5]引用了柳叶刀的论文[3]，针对武汉封城的有效性提出质疑，所说的原因是武汉封城只将对其它城市的扩散延后 2.9 天。我们认为这不是正确的推理。首先，从我们的模型预测，封城并不改变峰值的时间，而是感染人数的总量，另外，**不封城将造成对全国乃至全球更多的感染人群输出，会更快更强的加大病毒的传播**。

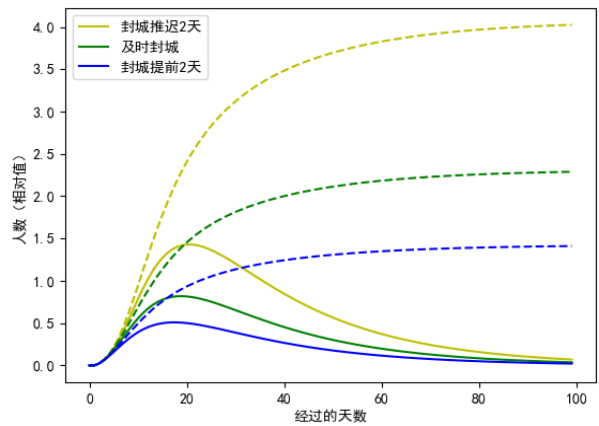

图 3.9. C-SEIR 模型预测对比：在不同时间点实施封城手段的疫情的影响

## 4. 总结

我们基于现有公布的数据，借助于传染病动力学模型，得到下列结论：
1. **武汉的封城举措，对于降低病毒感染人数具有重要的意义；**
2. **自政府采取相关防疫措施以来，全国各省市的病毒传染率均得到了较好的控制；**
3. **对疑似感染者的隔离观察是疫情防控的重要手段；**
4. **即使部分地区疫情似乎出现了拐点，但控疫思维和手段不可松懈，要避免二次高峰。**

我们的模型验证了隔离的重要性，但如何有效的隔离是一个需要进一步探讨的问题。严格来讲，隔离只是针对疑似感染人群。所以，如果检测设备足够灵敏便捷，人们的自我检测与隔离的意识足够强，理论上来讲，社会的工作和生活秩序可以基本恢复正常而不会影响疫情的变化。特别对武汉来说，**如果武汉的疑似感染人群都已被妥当的收治隔离，那么武汉的封城也就不再有疫情初期的意义了**。

从公共管理来讲，**如何进一步提高公共卫生水平，加强基础设施建设；如何针对公共交通和重点公共场所做有效的疏导，都值得相关政府部门提前筹划**。

疫情不同于病情，不仅关乎人们的身体健康，更是涉及到政治、经济、文化、教育，以及人们的心理健康与生活质量等各个方面。疫情防控手段与社会各要素之间的平衡关系恐怕需要一个更复杂的计算模型来评测；**如何在有效控疫与全社会利益之间找到一个平衡点，是一个更大的课题。**

由于篇幅有限，更多的数据和可视化会持续在 https://github.com/NCP-VIS 中更新，欢迎关注。

引用：
[1] 2019新型冠状病毒（2019-nCoV）疫情状况的时间序列数据仓库 https://github.com/BlankerL/DXY-2019-nCoV-Data
[2] 百度地图慧眼-百度迁移 http://qianxi.baidu.com/
[3] Joseph T Wu, Kathy Leung, Gabriel M Leung. Nowcasting and forecasting the potential domestic and international spread of the 2019-nCoV outbreak originating in Wuhan, China: a modelling study[J]. The Lancet (online first), January 31, 2020.
[4] Zhang J , Lou J , Ma Z , et al. A compartmental model for the analysis of SARS transmission patterns and outbreak control measures in China. Applied Mathematics and Computation, 2005, 162(2):909-924.
[5] Jon Cohen, Scientists are racing to model the next moves of a coronavirus that's still hard to predict. Science, Feb. 7, 2020.